\documentclass[11pt,a4paper]{article} 
\usepackage{psfig} 
\usepackage{epsfig} 
\usepackage{epsf} 
\usepackage{rotating} 
\usepackage{citesort}        
\usepackage{amsmath} 
\usepackage{amssymb} 
\usepackage{graphicx} 
\usepackage{latexsym} 
\def\MeV{{\rm MeV}} 
\def\eV{{\rm eV}} 

\def\npbps#1#2#3{  { Nucl. Phys. }(Proc. Suppl.){\bf B #1} (19#2) #3} 
\def\plb#1#2#3{    { Phys. Lett. }{\bf B #1} (19#2) #3}

\def\prl#1#2#3{    { Phys. Rev. Lett. }{\bf #1} (19#2) #3}

\begin{document} 
\begin{flushright} 
{\small 
FT-UM/04-123\\IFUM-871/FT\\ULB-04-167} 
\end{flushright} 
\vspace{0.2cm} 
\begin{center} 
{ \Large \bf  Analysis of Neutrino Oscillation Data with\\
         \bf  the Recent KamLAND results}\\[0.2cm] 
{\large 
P.~Aliani$^{\dagger}$\footnote{e-mails: 
paola.aliani@ulb.ac.be, vito.antonelli@mi.infn.it, 
ruggero.ferrari@mi.infn.it, 
marco.picariello@mi.infn.it, emilio.torrente-lujan@cern.ch }, 
V.~Antonelli$^{\star 1}$, R.~Ferrari$^{\star 1}$, 
M.~Picariello$^{\star 1}$, E.~Torrente-Lujan$^{\circledcirc 1}$\\[2mm] 
$^{\dagger}$ {\small\sl Service de Physique Th\'eorique, Univ. Libre
         de Bruxelles, Bruxelles, Belgium,}\\
$^{\star}$ {\small\sl Dip. di Fisica, Univ. di Milano}, 
{\small\sl and INFN Sez. Milano,  Via Celoria 16, Milano, Italy}\\ 
$^{\circledcirc}$ {\small\sl Dept. de Fisica, Grupo de Fisica Teorica, 
Univ. de Murcia, Murcia, Spain}} 
\end{center} 
 
\abstract{We present an updated analysis of all available solar and 
reactor neutrino data, emphasizing in particular the totality of the KamLAND 
results including the $SNO$ phase II (NaCl) spectrum data. 
In a two active-neutrino framework, we determine the solutions in the 
$\Delta m_{\odot}^2,\tan^2\theta_{\odot}$ parameter space compatible with 
experimental data using a $\phi$-distribution technique where the
gaussian approximation is not acceptable. 
Combining all data, we obtain the following best-fit parameters: 
$\Delta m_{\odot}^2 =8.17 \times 10^{-5} \eV^{2}$,
$\tan^2\theta_{\odot}=0.40$. The impact of these results is
discussed.}

\section{Introduction} 
Evidence of antineutrino disappearance in a beam of antineutrinos  
in the Kamland experiment has been presented \cite{klJun}. 
The analysis of the experimental results on reactor physics and solar 
neutrinos \cite{klothers} in terms of neutrino oscillations 
has largely improved our knowledge of neutrino mixing. 
\\ 
The recent KamLAND results confirm the previous data and give a 
more stringent limit on the neutrino mass parameters.  
\\ 
The solar neutrino data prior to September 7, 2003 converged to two 
distinct regions in the parameter space, often referred to as LMAI 
(centered around the best-fit point of
$\Delta m^2_{\odot}=7.1\times10^{-5} \eV^2$,
$\tan^2\theta_{\odot}=0.47$)
and LMAII (centered around
$\Delta m^2_{\odot}=1.5\times 10^{-4} \eV^2$, 
$\tan^2\theta_{\odot}=0.48$).
The inclusion of the last year's SNO data 
 eliminated the LMAII region at about $4\sigma$, and the new KamLAND
information further excludes the LMAII region ($\approx$ 5$\sigma$).  
The recent KamLAND results modify the best-fit point position in the
parameter space and the collaboration gives~\cite{klJun} a 
$\Delta m^2_{\odot}=8.2 \times 10^{-5}$ eV$^{2}$ and $\tan^2\theta = 0.40$.  
The aim of this work is to present a comprehensive updated analysis of 
all recent solar neutrino data including the KamLAND 
reactor-experiment results to determine the extent of the remaining 
viable region in the parameter space.  
\\ 
The structure of this paper is the following: in
section~{\bf\ref{kamland}} discuss our approache to the latest KamLAND
results. All solar neutrino experiments are discussed in
section~{\bf\ref{solar}}. We discuss the importance of the SNO data
and the spectrum results in section~{\bf\ref{NaCl}}. 
We then proceed, in section~{\bf\ref{analysis}}, 
to explain the procedure adopted in our 
analysis and in section~{\bf\ref{results}} we present our results. Finally 
we summarize and conclude in section~{\bf\ref{summary and conclusions}}. 
  
\section{KamLAND} 
\label{kamland} 
\subsection{The experimental setup} 
Reactor anti-neutrinos with energies above 1.8 MeV produced in some 53 
commercial reactors are detected in the KamLAND detector via the inverse 
$\beta$-decay reaction $\overline{\nu}_{e}+p\rightarrow n + e^{+}$. The mean 
reactor-detector distance and energy window of these $\overline{\nu}_{e}$ 
makes KamLAND an ideal testing ground for the LMA region of the $\nu_{\odot}$ 
parameter space. The first results published by the KamLAND collaboration 
eliminated all possible solutions to the solar neutrino problem (SNP) except 
the LMA region of the parameter space \cite{Eguchi:2002dm}. The 
sensitivity of this experiment to the $\Delta m^2$ parameter divided 
the previously whole LMA region into two distinct regions, the one 
relative to the smaller mass-squared difference being preferred by 
data \cite{postKamland}.   
\\ 
Recently, the KamLAND collaboration has announcent the new results: 
they give the results of the data taken from October to Jan 2004, 
corresponding to about 317 days live time. The collaboration has taken all 
data and in the analysis, has included the previous set. The most
significant changes in the analysis technique are related to the
fiducial volume definition.
Whereas in the previous work, events taking place at the outer edge of
the nylon balloon were rejected, the recent analysis adopts a more
sophisticated coincidence-measurement technique to exclude unwanted
backgrounds. 
The estimated error on initial $\phi_0\left(\overline{\nu}_{e}\right)$
is of 2\%, and
a discussions between scientists and representatives of the 
commertial reactors has allowed the KamLAND collaboration to estimate
the incoming neutrino
flux with better accuracy. A graph of the variation of initial flux
throughout the period of
data taking is shown in fig. 1.a of \cite{klJun}.  
\subsubsection{Our simulations}\label{simulations} 
In order to correctly model the detector and incorporate it into our 
solar neutrino analysis, we used a constant, time-averaged fuel 
composition for all of the commercial reactors within detectable 
distance of the Kamioka site, namely  ${}^{235}\text{U}=56.3\% $, 
${}^{238}\text{U}=7.9\% $, ${}^{239}\text{Pu}=30.1\% $, and 
${}^{241}\text{Pu}=5.7\% $. We used the full cross-section including 
electron recoil corrections. We analyzed the data above   
threshold of $2.6 \MeV$, as the low-energy end of the 
spectrum had effectively no events. We exclude, for the same reasons,
the last bin.
We neglected all backgrounds in 
our analysis, including geological background above 2.6 MeV.  
We use the resolutions published by the collaboration for the two different 
data sets, namely $\sigma\left(E\right) = 6.2\%/\sqrt{E}$ for the recent data  
and $\sigma\left(E\right) = 7.3\%/\sqrt{E}$ for data previousely published. 
\\ 
In order to use all the data available, we use our montecarlo to estimate an 
equivalent efficiency for the two pre-upgrade and post-upgrade
phases. In this way, we can
determine an effective efficiency which takes the upgrade into
account. The information relative
to the no-oscillation initial flux for the two periods can be
extracted from fig. 1.a of \cite{klJun}.
In their report, the collaboration gives an energy resolution
$\sigma/E$ of about $6.2 \%$ at $1 \MeV$, and bins ranging fro 1.8 MeV
to 8.5 MeV, althought the lower energy bins are more affected by
geo-neutrino backgrounds and are excluded from the analysis. The total
systematic error is estimated at $6.5 \%$. The spectrum can be deduced
from fig. 2.a of \cite{klJun} and is summerized in table~[\ref{t2}]. 
\\ 
The collaboration adopts a hybrid statistical technique to account for
the non-gaussian
high energy bins, where the number of events are insufficient. The
measure suited to samples from a multinomial
distribution, the Pearson-$\chi^2$, can be generalized to work with
samples from arbitrary asymptotic multivariate
distributions. This corresponds to the transition from independent an
system to a correlated one. Becaouse we
want to take into account the correlation between the systematic
errors, we do not use the Pearson-$\chi^2$, but 
its generalization, the $\phi$-distribution in order to take
systematic effects into account. Furthermore, we 
assume full correlation between the different bins. 
\\ 
No MSW effects are taken into consideration for the KamLAND data
alone, as it was shown that
any asymmetry due to matter effects is negligible for $\Delta m^2$ of
the order of $10 \times 10^{-5}$ eV$^2$.
 
\section{Solar data} 
\label{solar} 
The most ponderous data present in our analysis come from the solar 
neutrino experiments. The experimental results are compared to an 
expected signal which we compute numerically by convoluting the solar 
neutrino fluxes \cite{bpb2001},  oscillation probability at the 
detector location and detector response functions. Details of the 
numerical techniques used in this analysis can be found elsewhere 
\cite{ourstuff}. Double-binned day-night and zenith angle bins are 
computed in order to analyze the full SuperKamiokande data 
\cite{Smy:2002fs}, whereas single-binned data is used for the SNO 
detector \cite{newSNO,Poon:2001ee,Ahmad:2002jz}. The global signals
only are used for the
radiochemical experiments Homestake \cite{homestake}, SAGE 
\cite{sage1999,sage}, GallEx \cite{gallex} and GNO \cite{gno2000}. 
In particular, we want to draw attention to the fact that we have used 
both the global {\it and} the spectrum data of the phase-II SNO results. 
 
\subsection{The Sudbury Neutrino Observatory} 
\label{NaCl} 
The Sudbury Neutrino Observatory (SNO) collaboration has 
presented data relative to the NaCl phase of the experiment 
\cite{newSNO}. The addition of  
NaCl to a pure $\text{D}_{2}\text{O}$ detection medium has the effect 
of increasing the detector's sensitivity to the neutral-current (NC) 
reactions within its fiducial volume. The NC detection efficiency has 
changed from a previous 'no-salt' phase of approximately a factor 
three (the efficiency being a position-related quantity, we see that 
the increase is evens is bigger for distant events (see fig. 1a of 
\cite{newSNO}). This novelty has made it possible for the SNO 
collaboration to analyze their data without making use of the 
no-spectrum-distortion hypothesis. Furthermore, they have adopted a
new 'event
topology' criterion to distinguish among the different channels within 
the detector.  
\\ 
The SNO detector was planned to run on a three-phase program (for a 
comprehensive coverage of the SNO potentiality, see 
\cite{Boger:1999bb}). Being a heavy-water Cherenkov detector, it is 
sensitive to three neutrino interactions: elastic scattering (ES) 
$\nu_{w}+e^{-}\rightarrow \nu_{x}+e^{-}$, charged-current (CC) 
$\nu_{e}+d\rightarrow e^{-}+2p$ and neutral current (NC) 
$\nu_{x}+d\rightarrow p+n+\nu_{x}$. During the first phase of 
operation, data relative only to the first two of these was collected 
\cite{Poon:2001ee}. Successively, the NC reaction was tagged 
\cite{Ahmad:2002jz}. This reaction, which is equally sensitive to all 
neutrino flavors, is a fundamental experimental result, as it not only 
gives a direct measurement of the $^8$B flux, but also represents a 
direct measurement of non-electron-flavor neutrinos originating in the 
sun. Unfortunately, although the second-phase results proved essential 
in confirming the oscillation hypothesis, the efficiency with which the 
NC reactions were detected at SNO was quite low. The results depended 
heavily on the explicit assumption that the neutrino spectrum be 
non-distorted, an assumption that is fortunately true in the LMA 
region of the parameter space although misleading in the MSW 
regime. It was perhaps this observation that pushed certain authors to 
analyze the effects of varying the weight of the matter interaction 
term in the neutrino propagation Hamiltonian \cite{fogli-lisi}. In 
this context, it was shown  that despite the extraordinary amount of 
solar neutrino data in our possession, if a fit to different 'weights' 
of the matter interaction term did not prefer in a significant way the 
classical MSW solution, and in fact, the data did not exclude the 
no-MSW effect hypothesis \cite{fogli-lisi}. This analysis has been 
carried out again with the new SNO data. The inclusion of this data 
seems to exclude the null hypothesis at $5.6\sigma$ \cite{new_fogli 
lisi}.  
\\ 
The addition of NaCl to the pure heavy water in the SNO detector has
the effect of
drastically increasing the detector's sensitivity to the NC 
reaction. The $^{35}$Cl acting as receptors for the neutrons emitted 
in the NC reaction, produce a cascade of photons  which are easier to 
detect. The NC signal was tagged as being a 'multiple-$\gamma$' event 
as opposed to the 'single-$\gamma$' CC and ES signals in this new data 
set.  
\\ 
The SNO Collaboration has now devised a new data-analysis technique 
which relies on the topology of the three different events. The new 
parameter ($\beta_{\ell}$) relative to which they marginalize is known 
as the 'isotropy' of the Cherenkov light distribution was used to 
separate the CC, ES and NC signals, something that was not possible in 
th previous two data sets. The measured fluxes as reported in 
\cite{newSNO} are:  
\begin{eqnarray} 
\phi_{\text{{\tiny CC}}}^{\text{{\tiny SNO}}} & = &
 1.59^{+0.08}_{-0.07}\text{ (stat) }^{+0.06}_{-0.08}\text{ (syst) }
\nonumber \\
\phi_{\text{{\tiny ES}}}^{\text{{\tiny SNO}}} & = &
2.21^{+0.31}_{-0.26}\text{ (stat) }\pm 0.10\text{ (syst) } \\
\phi_{\text{{\tiny NC}}}^{\text{{\tiny SNO}}} & = & 5.21\pm 0.27
\text{ (stat) }\pm 0.38\text{ (syst) }\nonumber \\
\end{eqnarray} 
yielding a CC/NC ratio of 
\begin{equation} 
\phi_{\text{CC}}^{\text{SNO}}/\phi_{\text{NC}}^{\text{SNO}}=0.306\pm
 0.026\text{ (stat) }\pm 0.38 \text{ (syst)}
\end{equation} 
For the purpose of their analyzes, the SNO collaboration did not make 
use of the spectral information to produce their contour plots (see 
the SNO HOWTO for details \cite{howto}), both for simplicity and 
scarcity of information contained in the data itself. Nonetheless, we 
make use of all published data and refer the reader to
section~{\bf\ref{analysis}} for details.
\\ 
The comparison of the new SNO results and the previous phase-II data 
can easily be made because the SNO collaboration has included in the 
recent paper results which were obtained following their previous 
method, along with the new unconstrained data. The new results are 
compatible with the previous ones. It seems that the overall effect of 
un-constraining the analysis is an increase in the measured fluxes, 
although the estimated total $\Phi_{\text{B}}$ has decreased relative 
to the previous best-fit value, leaving even less space for eventual 
sterile neutrinos.  
 
\section{Our analysis} 
\label{analysis} 
We use a standard $\chi^2$ technique to test the non-oscillation 
hypothesis. Two different sets of analyses are possible with the 
present data on neutrino oscillations:  
1) short-baseline reactor data, solar data including the SK spectrum  
and previous phase-I (CC only) SNO spectrum, phase-II SNO global
result, combined with new the KamLAND spectrum and, 
2) the previous set with the use of the phase-II SNO spectrum result
and the new KamLAND data.
In order to use all the SNO data, we consider the phase-I and phase-II
results as two distinct but fully correlated experiments. 
\\ 
In all cases, the $\chi^{2}$ for the global rates of the radiochemical 
experiments are  
\begin{equation}\label{chi_radiochemical} 
\chi^{2}_{\text{rad}}=\left(\mathbf{R}^{\text{th}}-\mathbf{R}^{\text{exp}}\right)^{T}\left(\sigma_{\text{corr}}+\sigma_{\text{uncor}}\right)^{-1}\left(\mathbf{R}^{\text{th}}-\mathbf{R}^{\text{exp}}\right),
\end{equation} 
where $\mathbf{R}^{\text{th,exp}}$ are length-two vectors containing the  
theoretical (or experimental) signal-to-no-oscillation expectation for 
the chlorine and gallium experiments. Correlated systematic and uncorrelated 
statistical errors are considered in $\sigma_{\text{syst}}$ and 
$\sigma_{\text{stat}}$ respectively. Note that 
the parameter-dependent $R^{\text{th}}$ is an averaged day-night quantity, as 
the radiochemical experiments are not sensitive to day-night variations. 
\\ 
We consider the double-binned SK spectrum comprising of 8 energy bins for a 
total of 6 night bins and one day bin. The $\chi^{2}$ is given by 
\begin{equation}\label{chi_spectrum_sk} 
\chi^2_{\text{SK}}= (\alpha\mathbf{R}^{\text{th}} -\mathbf{R}^{\text{exp}})^T  
\left (\sigma^{2}_{\text{unc}} + \sigma^{2}_{\text{cor}}\right )^{-1} 
 (\alpha\mathbf{R}^{\text{th}}-\mathbf{R}^{\text{exp}}). 
\end{equation} 
The covariance matrix $\sigma$ is a 4-rank tensor containing 
information relative to the statistical errors and energy and 
zenith-angle bin-correlated and uncorrelated uncertainties. 
Since the publication of the first SNO NC results, we have adopted 
their estimate of $\phi_{B}$ and incorporated the new parameter 
$\alpha$ in the $\chi^2$ representing the normalization with respect 
to this quantity. In determining our best-fit points, we minimize with 
respect to it. 
\\ 
As discussed in section~{\bf\ref{simulations}}, 
 we use an alternative technique 
for the KamLAND data due to the low statistics of the high energy
bins. Furthermore, in order to use all the available data,  
we determine overall detector efficiencies for  
the pre- and post- PMT/electronics upgrade phases and use them
subsequently in our analysis.
\\ 
The total KamLAND contribution to the $\chi^2$ is therefore: 
\begin{equation}\label{chi_kamland} 
\chi^2_{\text{KL}} = \chi^2_{\text{KL,glob}}+\chi^2_{\text{KL},\phi} 
\end{equation} 
where 
\begin{equation}\label{chi_kamland_both} 
 \chi^2_{\text{KL,glob}}= 
 \frac{\left(R^{\text{th}}-R^{\text{exp}}\right)^2}{\sigma_{\text{stat+sys}}^2}
\end{equation} 
In the evaluation of $\chi^2_{\text{KL},\phi}$ we use vectors that
comprise therefore of 13 spectral points of width 0.425 MeV.  
We have studied the influence of the systematic errors on the KamLAND
response function and
incorporated them in our correlation matrix. We assume full correlation among 
the different bins. Note that the quantities $\mathbf{R}^{\text{exp}}$ and 
$\mathbf{R}^{\text{th}}$ contain the number of events normalized to the 
no-oscillation scenario. Due to the fact that at high energy KamLAND
observes a small number
of events, we cannot use a Gaussian approximation for all bins. This
means that the  correlated  
systematic deviations cannot be introduced in a straightforward way. 
The $\phi$-distribution is used for those bins and refer the reader to 
the litterature for discussions on the applicability of this 
method~\cite{stat}. 
 
\subsection{Global SNO signal and other neutrino oscillation data} 
We recall that two analyzes of the SNO data are presented. 
The first consideres the phase-II  
global signal alone, the second incorporates the phase-II spectrum. 
\\ 
For the purpose of this analysis, the $\chi^{2}$ can be considered as
the sum of three distinct contributions: 
\begin{equation} 
\chi^{2}=\chi_{\text{KL}}^{2}+\chi_{\odot/\text{SNO}}^{2}+\chi^2_{\text{glob,SNO}}
\end{equation} 
where $\chi^2_{\text{KL}}$ is as in eq.~\ref{chi_kamland_both}. 
\\ 
We define $\chi^2_{\odot/\text{SNO}}$ as the contribution of all solar
neutrino experiments except SNO, i.e.
(see eq.~\ref{chi_radiochemical},~\ref{chi_kamland_both}):
\begin{equation}\label{chi_solar_noSNO} 
\chi^2_{\odot/\text{SNO}}=\chi^2_{\text{rad}}+\chi^{2}_{\text{SK}} 
\end{equation} 
The SNO contribution is given by 
\begin{equation}\label{chi_SNO} 
\chi^{2}_{\text{glob,SNO}}=\left(\mathbf{R}^{\text{th}}-\mathbf{R}^{\text{exp}}\right)^{T}\left(\sigma_{\text{corr}}+\sigma_{\text{syst}}\right)^{-1}\left(\mathbf{R}^{\text{th}}-\mathbf{R}^{\text{exp}}\right) 
\end{equation} 
where $\mathbf{R}^{\text{th,exp}}$ are length-14 vectors containing 
the phase-II spectrum data (13 bins) and the new phase-III global SNO 
result. We consider the two SNO results as if coming from two 
independent experiments, but fully correlated. We use the backgrounds 
as listed in table 1 of \cite{Ahmad:2002jz} for the phase-II data, 
and the detector resultion given in the HOWTO. 
 
\subsection{The SNO spectrum and other neutrino oscillation data} 
\label{spectrum} 
As mentioned previously, the SNO collaboration does not analyze the full 
spectrum data. We prefer, instead, to consider this information as well. 
We therefore consider the full SNO phase-II and phase-III data (listed 
in table~[\ref{sno_spectrum}]). The $\chi^{2}_{\text{spect,SNO}}$ has 
the same formal experssion as eq.~\ref{chi_SNO}, where it is 
understood that $\mathbf{R}^{\text{th,exp}}$ are now length 26 
containing two 13-bin relative to the two SNO data 
sets. We consider the two fully correlated. 
\\ 
The difficulty in using the spectrum data lies in correctly estimating 
the systematics. By using the information contained in table II of 
\cite{newSNO}, we have computed the influence of all the different 
sources of error on our response function considering the 
correlation/anti-correlation as presented in table 1 of 
\cite{howto}. The different backgrounds present yet another 
difficulty, because we have access only to the total number of 
background events coming from the different sources listed in table I 
of \cite{newSNO}. We have taken the background spectrum presented in 
the phase-II paper and normalized it to the total number of events in 
the $\text{D}_{2}\text{O + NaCl}$ phase. 

\section{Our results} 
\label{results} 
\begin{figure} 
\centering 
\begin{tabular}{cc} 
\includegraphics[scale = 0.7]{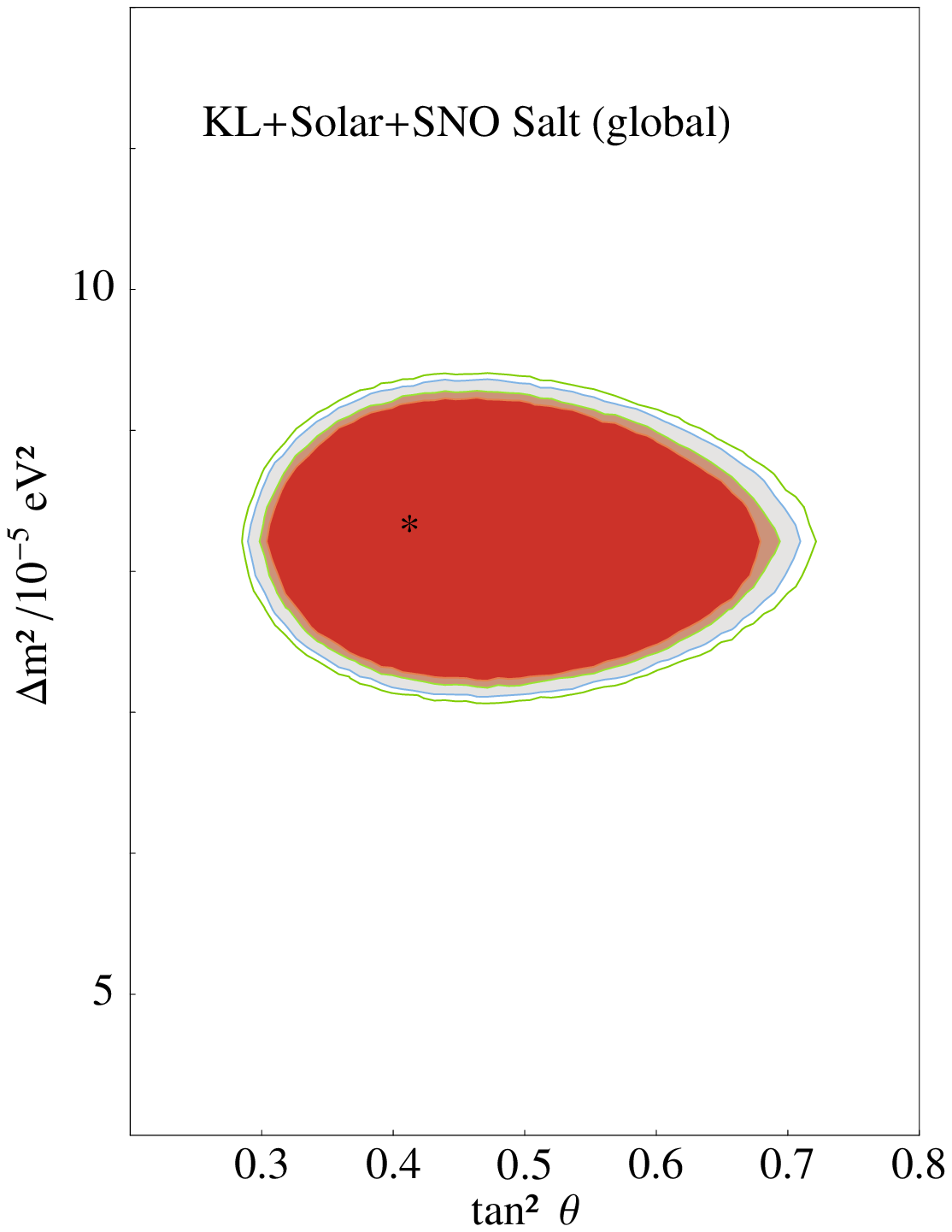}& 
\includegraphics[scale = 0.7]{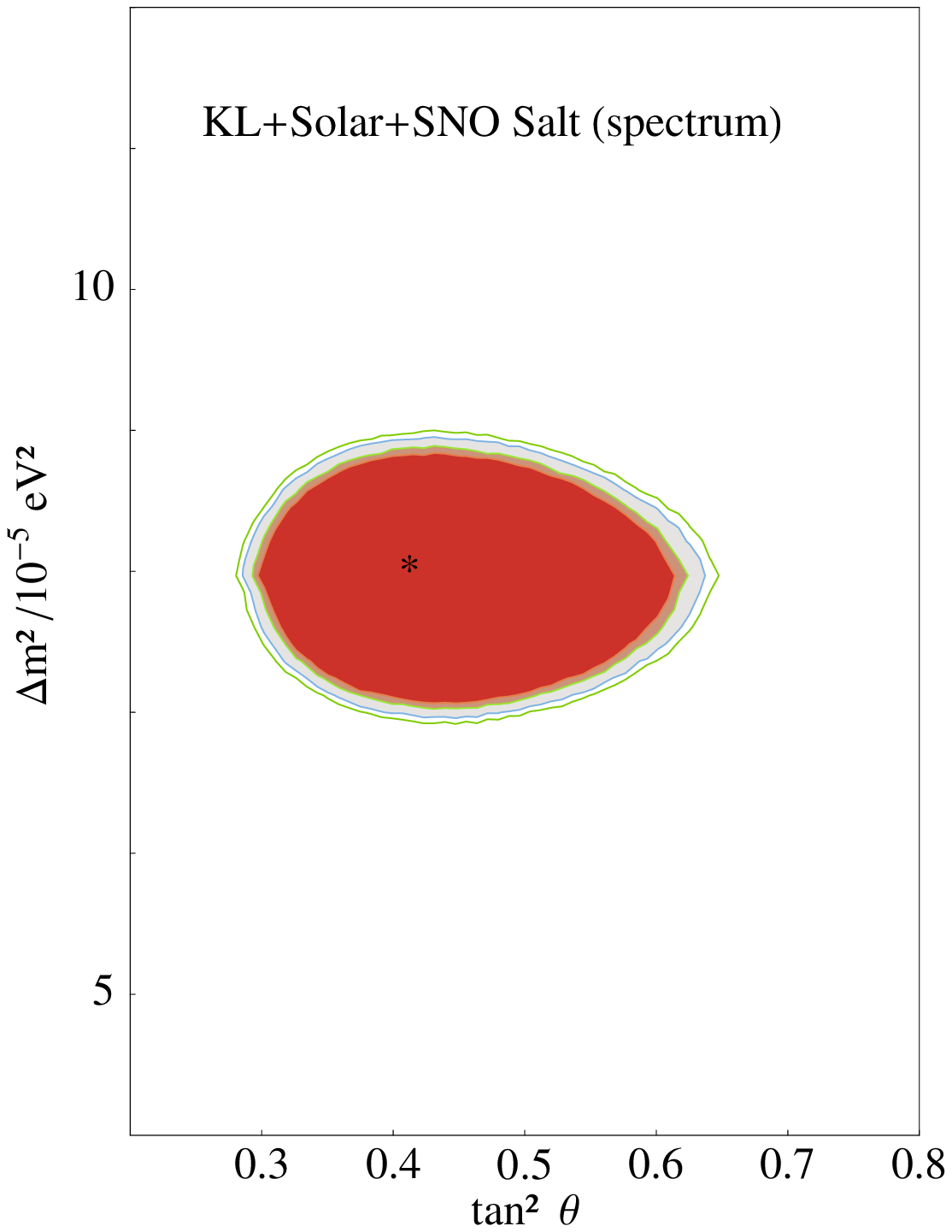} 
\end{tabular} 
\caption{\small   
({\it left}) Allowed region in the $(\tan^2\theta, \Delta
m_{\odot}^2)$ plane for the global analysis, which includes the
previous solar data (see e.g. \cite{ourstuff} for details) and all KamLAND Global results.  
({\it right}) Best fit solution for the spectrum analysis, including
all previous solar data,  short baseline reactor data and the KamLAND
spectrum. Best fit point given in table [\ref{bestfitpoints}].} 
\label{exclusions} 
\end{figure}

In table~[\ref{bestfitpoints}] we report the values of the mixing parameters 
$\Delta m^2_{\odot}$, $\tan^2\theta_{\odot}$, and the $\chi^2$ 
obtained from minimization and from the peak of marginal likelihood 
distribution. 
\\ 
In fig.~[\ref{exclusions}-(left)] we show the exclusion plots for the
solar, radiochemical + Cherenkov solar data and KamLAND with the
global signal of the SNO phase-II data, whereas the right 
pannel refers to the KamLAND spectrum, radiochemical + cherenkov solar data 
and the SNO phase-II spectrum information. Contour lines correspond to the 
the allowed areas at 90, 95, 99 and 99.7\% CL 
relative to the absolute minimum.  
\\ 
In fig.~[\ref{marginalization}] 
we plot the marginalized likelihood distributions for 
each of the oscillation parameters $\Delta m^2$, and $\tan^2 \theta$ 
corresponding to the Kamland Spectrum plus  solar evidence. 
The values for the peak positions and their widths are obtained by fitting 
two-sided Gaussian distributions and are given in 
table~[\ref{bestfitpoints}.a]. 
 
\section{Summary and  Conclusions}\label{summary and conclusions} 
\begin{figure} 
\centering 
\begin{tabular}{cc} 
\includegraphics[scale = 0.7]{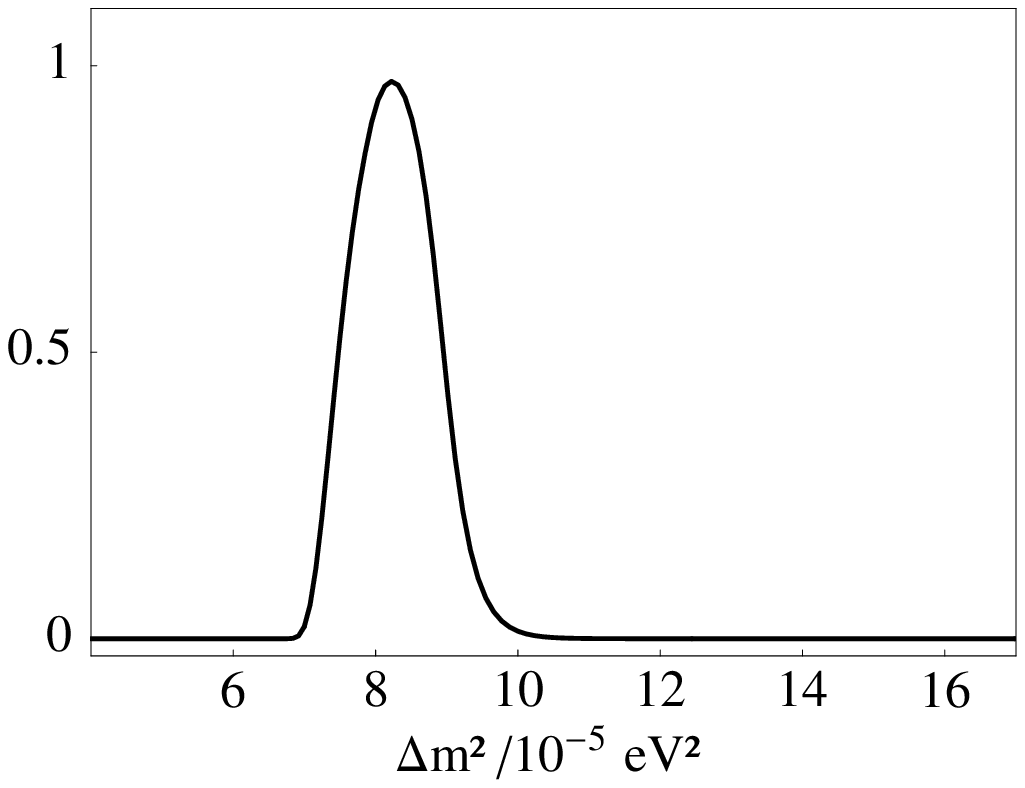}& 
\includegraphics[scale = 0.7]{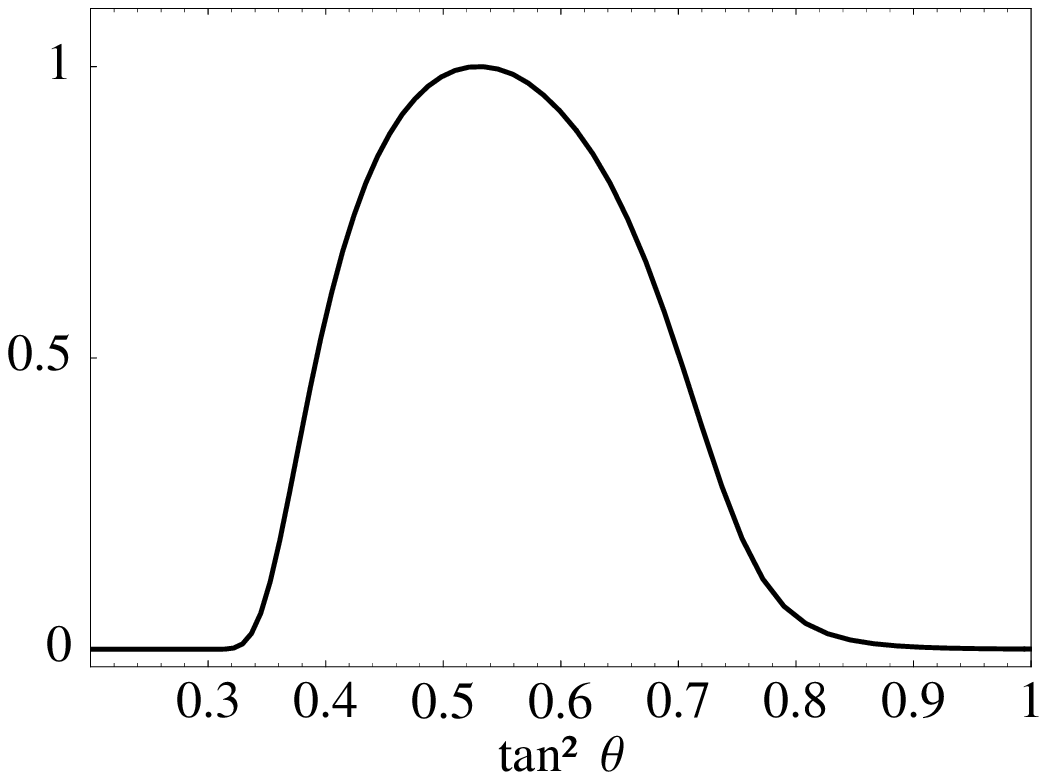} 
\end{tabular}\caption{\small   
marginalized likelihood distributions for each of the 
oscillation parameters $\Delta m^{2}_{\odot}$ (left) and  
$\tan^2\theta$ (right) corresponding to the totality 
of solar and KamLAND data. The curves are in arbitrary 
units with normalization to the maximum height. 
} 
\label{marginalization} 
\end{figure}

We have presented an up-to-date  analysis including the recent kamland
results, the SNO-phase II spectrum and all other solar neutrino data
The (predominantly) active  
neutrino oscillations hypothesis has been confirmed, and the
decoupling of the atmospheric $\Delta m^2$-solar 
$\Delta m^2$ justifies a 2-flavour analysis. 
\\ 
Due to the increased statistics, the inclusion of the new KamLAND data
determines with good accuracy the value of $\Delta m^2_{\odot}$, clearly 
selecting te LMAI solution, and 
brings us to a new era of precision measurements in the solar neutrino
parameter space \cite{kamlandTalk}.
\\ 
It is interesting to note that the KamLAND data alone
predicts, for both their analyses, a value of tan2theta smaller than
the one obtained with the previous data, and significantly different
from 1, consequently making the aesthetically pleasing bi-maximal-mixing
models strongly disfavored. This result confirms what was already evident
in the solar neutrino data analyses. Nevertheless, improvement on the
determination of $\tan^2\theta$ is necessary and it is known that KamLAND is  
only slightly sensitive to this mixing parameter. The (lower) accuracy
with which we determine the solar mixing angle is evident in the marginalized 
likelyhood plots of fig.~[\ref{marginalization}].  
Planning of future super-beam experiments
aimed at determining the $\theta_{13}$ and eventual CP violating phases
relies on the most 
accurate estimation of all the mixing parameters~\cite{terranova}. 
It is expected that future solar neutrino experiments, notably phase-II SNO
(higher statistics, due to be made public soon)
and eventually future low energy experiments, and phase-III
SNO (with helium) will further restric the allowed range of parameters.
\\ 
The use of the SNO phase-II spectrum in the data set has mainly two
effects: 1) a slight reduction in the overall area of the exclusion plot
and 2) a slight decrease in the best-fit $\Delta m^2_{\odot}$.
\\The decrease in the
best-fit mass squared difference can be understood by the fact that by
including the SNO spectrum, we increase the statistical relevance of solar
neutrino data, which {\em prefer smaller $\delta m^2$}. Furthermore, the
oscillation pattern (whose information is contained in the spectrum) is
more sensitive to $\Delta m^2$.

\vskip 1truecm 
\noindent{\bf\large Acknowledgments}\\ 
We would like to thank F. Terranova and M. Smy for usefull discussions. 
P.A and V.A. thank the brasserie/patisserie in Rue Jean de Beauvais 
for providing a 
suitable working environment in Paris. The computations were done at the  
computer farm of the Universit\`a degli Studi di Milano, Italy.

\newpage 
 
\begin{table}[p] 
\begin{center} 
\begin{tabular}{cc|cc} 
\hline\hline 
$T_{eff}$ (MeV)& Events per 500 keV &$T_{eff}$ (MeV)& Events per 500 keV\\ 
\hline 
 5.5- 6.0    & 510 & 9.0- 9.5    & 130\\ 
 6.0- 6.5    & 515 & 9.5-10.0    & 120\\ 
 6.5- 7.0    & 475 &10.0-10.5    &  60\\ 
 7.0- 7.5    & 385 &10.5-11.0    &  70\\ 
 7.5- 8.0    & 300 &11.0-11.5    &  40\\ 
 8.0- 8.5    & 240 &11.5-12.0    &  35\\ 
 8.5- 9.0    & 145\\ 
\hline 
\multicolumn{4}{l}{{\tiny $^{\dagger}$ It contains the CC, ES, and NC 
+ internal and external source neutron events.}}\\ 
\hline\hline 
\end{tabular} 
\end{center} 
\caption{{\small Energy spectrum observed at SNO (taken from fig.~2 of 
ref~\protect{\cite{newSNO}}).$^{\dagger}$}}\label{sno_spectrum} 
\end{table} 

\begin{table} 
\begin{center} 
\label{results table} 
\begin{tabular}{lcr} 
\hline\hline 
{\large {\bf Experiment}} &&{\large {\bf References}}\\ 
\hline 
Homestake& &\cite{homestake}\\ 
SAGE & &\cite{sage1999,sage}\\ 
GallEx &&\cite{gallex}\\ 
GNO &&\cite{gno2000}\\ 
\\ 
SuperKamiokande & &\cite{Smy:2002fs}\\ 
SNO & & \cite{newSNO,Poon:2001ee,Ahmad:2002jz}\\ 
\\ 
CHOOZ && \cite{chooznew} \\ 
Palo Verde & & \cite{PaloVerde}\\ 
KamLAND & &\cite{klJun,Eguchi:2002dm}\\ 
\hline\hline 
\hline 
\end{tabular} 
\caption{References from where we draw the data used in our analysis} 
\end{center} 
\end{table} 
\clearpage 
 
\begin{table}[p] 
\begin{center} 
\begin{tabular}{lll} 
\hline\hline 
 & $\Delta m^2 (\eV^2) $& $\tan^2\theta$\\ 
\hline 
\multicolumn{3}{l}{from $\chi^2$ minimization}\\[0.17cm] 
\hspace{0.3cm} KL (Sp+Gl)+Solar + SNO (Sp)&  $7.89\times 10^{-5}$ & $0.40$ \\[0.15cm] 
\hspace{0.3cm} KL (Sp+Gl)+Solar + SNO (Gl)&  $8.17\times 10^{-5}$ & $0.40$ \\[0.25cm] 
\multicolumn{3}{l}{from marginalization}\\[0.17cm] 
\hspace{0.3cm} KL (Sp+Gl)+Solar + SNO (Sp)&  $8.2^{+0.9}_{-0.8}\times 10^{-5}$ & $0.50^{+0.12}_{-0.6}$ \\[0.25cm] 
\hline 
\end{tabular} 
\caption{\small Mixing parameters  
 from $\chi^2$ minimization and likelihood marginalization. 
}\label{bestfitpoints} 
\end{center} 
\end{table} 
 
\begin{table} 
\centering 
\begin{tabular}{lcll} 
\hline \hline 
Bin (MeV) &   S$_{\text{exp}}$/MC & $\sigma_{\text{stat.}}$ (\%) \\ 
\hline 
2.600 - 3.025 & 0.45 & 1.4\\ 
3.025 - 3.450 & 0.56 & 1.5\\ 
3.450 - 3.875 & 0.67 & 1.7\\ 
3.875 - 4.300 & 0.62 & 2.1\\ 
4.300 - 4.725 & 0.99 & 2.6\\ 
4.725 - 5.150 & 1.20 & 3.3\\ 
5.150 - 5.575 & 0.80 & 4.5\\ 
5.575 - 6.000 & 1.00 & 6.7\\ 
6.000 - 6.425 & 1.20 & 10.0\\ 
6.425 - 6.850 & 0.33 & 17.0\\ 
6.850 - 7.275 & 0.67 & 33.0\\ 
7.275 - 7.700 & 0.00 & - \\ 
7.700 - 8.125 &  -   & - \\ 
 \hline\\ 
\end{tabular} 
\caption{Summary of Kamland spectrum information extracted  
from \cite{klJun}. Relative statystical errors only are reported.} 
\label{t2} 
\end{table} 

\clearpage

\end{document}